\documentclass[aps,superscriptaddress,showpacs,twocolumn]{revtex4}
\usepackage[dvips]{graphics}

\begin{document}
\title{Ergodic transitions in continuous-time random walks}

\author{Alberto Saa}\email{asaa@ime.unicamp.br}
\affiliation{Departamento de Matem\'atica Aplicada, UNICAMP,  13083-859 Campinas, SP, Brazil}
\author{Roberto Venegeroles}\email{roberto.venegeroles@ufabc.edu.br}
\affiliation{Centro de Matem\'atica, Computa\c c\~ao e Cogni\c c\~ao, UFABC, 09210-170 Santo Andr\'e, SP, Brazil}

\date{\today}

\begin{abstract}
We consider    continuous-time random walk models
 described by   arbitrary sojourn  time probability density functions.
We find a general expression
 for the distribution of time-averaged observables for such systems,
 generalizing some recent results presented in the literature.
 For the case where sojourn times  are identically distributed independent random variables, our results shed some light on the recently proposed
  transitions between ergodic and weakly nonergodic regimes. On the other hand, for the case of non-identical
 trapping time  densities over the lattice points,
 the distribution of time-averaged observables reveals
 that such systems are typically nonergodic, in
  agreement with some
 recent experimental evidences on the statistics of blinking quantum dots.
 Some explicit examples are considered in detail. Our results are independent
 of the lattice topology and dimensionality.
\end{abstract}


\maketitle

\section{Introduction}
\label{sec1}
Boltzmann's hypothesis of ergodicity is a central concept in Statistical
Mechanics. Roughly speaking, in an ergodic system, for a long time observation,
 the residence time of a trajectory in a given region of the phase space
 is proportional  to the volume measure of the region.
 Despite the high success of Boltzmann's description of large systems, the
 ergodic
 hypothesis cannot be used, for instance, for systems whose phase space can be subdivided in  mutually inaccessible regions. A subtler physical scenario  of nonergodicity was introduced by Bouchaud \cite{Bou} in the context of glass dynamics: the so-called {\it weakly nonergodic} systems are also nonergodic, but their phase spaces are not subdivided
  in  mutually inaccessible regions. Recently,
  there has been great interest \cite{RB,BB,HBMB},
  in particular, in
  the weak ergodicity breaking phenomenon, where
  a transition to an ergodic phase may occur.  Notice that
Bouchaud's   ideas   about weak ergodicity breaking are, in turn, closely related to concepts that have been previously considered  in the mathematical literature about ergodic theory and stochastic processes, see, for instance \cite{Aaronson}. In particular, a weakly nonergodic regime corresponds to
a situation in which the state space of a (semi)-Markov process is connected,
 {\em i.e.},
any state can be reached from any other state with finite probability in a finite number of
steps, but the fraction of occupation time for a given state is not equal to its invariant spatial
measure.

From the physical point of view, weakly nonergodic systems have proved to be relevant in many applications as, for instance,
complex networks \cite{WGG}, weak turbulence \cite{SFGA}, and, in
particular, they
  are  at the basis for the statistical modeling of
   atoms trapping by laser cooling devices \cite{BBAT,SLCT}. In these models,
 the atom quantum dynamics   are equivalent to a classical  random walk
  in the momentum space, where the standard deviation of the jump lenghts $\Delta p$ is of the same order of the incident photon momentum. The
  trapping
  process  consists effectively in successive frontal collisions,
  for some given time interval, between the atom and
   the resonant laser photons,
   in a process   called   subrecoil cooling. Statistically, it
   is equivalent
   to the application of a controllable potential jump rate $R(p)$,
    responsible for the trapping near $p=0$. The resulting scenario is a kind of continuous-time
    random walk where rare events play a dominant role, with
    L\'evy type
     non-gaussian probability density functions (PDF)   $\psi(\tau)\sim A_{\alpha}\tau^{-(1+\alpha)}$
      governing  the trapping times.
The divergence of the average trapping time
\begin{equation}
\label{bartau}
\bar{\tau} = E  \left(  \tau\right) = \int_0^\infty \tau \psi(\tau) \, d\tau
\end{equation}
 for  $0<\alpha<1$, with
   $E \left( \ \right)$
   denoting the expectation value with respect to the PDF $\psi(\tau)$,
  is precisely at
 the root  of the subrecoil cooling mechanism effectiveness.
Such mechanisms are typically nonergodic, and the origin
 of the nonergodicity is usually attributed  to the divergence of the average trapping time \cite{BBAT}.

We investigate here the PDF of time-averaged observables for a general class of continuous-time random walk models (CTRW), extending considerably the classes already considered in
 \cite{RB}. The nonergodic  properties of CTRW models are studied here by
 comparing PDF of ensemble averages with fluctuations of time averages.
 Initially, we consider   CTRW models described by an arbitrary trapping time PDF $\psi(\tau)$, in which all the sojourn times $\tau$ of the $L$ lattice points $1\leq x\leq L$ are identically distributed independent random variables.
 We show that models with finite $\bar\tau$ are always ergodic.
In the cases where
 $\bar\tau$ diverges mildly, one still has ergodicity, but  for stronger    divergences the
 dynamical regime is weakly nonergodic and characterized by a Lamperti type PDF  of time-averaged observables, in agreement with the recent results announced
 in \cite{RB}.
   We also extend our approach to include CTRW models with non identical trapping time PDF over the lattice points. For such cases, we show that even models with
 $\bar{\tau}$   finite for all lattice points are typically
  nonergodic. The obtained PDF of time-averaged observables clarify
 some  nonergodic features   reported recently for blinking quantum dots systems \cite{BHMDBD,MB}.

The paper is organized as follows. In Section \ref{sec2}, we provide the basic concepts and definitions required to describe ergodic and nonergodic properties of CTRW models. In Section \ref{sec3}, the
distribution density of time-averaged observables
  for the CTRW model with  identical trapping time PDF is obtained.
    Section \ref{sec4} is devoted to discuss the   weak ergodicity breaking
    phenomenon, extending some of the results obtained in \cite{RB}.
     Some explicit examples of PDF are considered in Section \ref{sec5}.
     In Section \ref{sec6}, we investigate the CTRW with non identical
     trapping time PDF.
     We conclude in Section \ref{sec7} with some remarks about possible universal aspect of nonergodic fluctuations close to the weak ergodicity breaking.

\section{Statistical approach for CTRW models}
\label{sec2}

Here, we consider a general   CTRW model on a  lattice with   points $x=1,\ldots,L$, subject to certain trapping statistics.
The topology and the dimensionality of the lattice are irrelevant for
our purposes.
The notion of random waiting time between successive steps was originally introduced by Montrol and Weiss \cite{MW}. In the model discussed here, the particle can jump to one of its nearest neighbors after waiting for a random time $\tau$. The trapping statistics is given by the waiting time $t_{x} = \sum_{i=1}^{n_{x}}\tau_{i}$, where $\tau_{i}$ is the $i$th sojourn time of a given
lattice site $x$. The set $\left\{\tau_{1},\ldots,\tau_{n_{x}}\right\}$ for $1\leq x\leq L$ is composed of non-negative and identically distributed independent random variables with a common and arbitrary PDF $\psi(\tau)$. As in Ref. \cite{RB}, we also consider here $n_{x}=nP_{x}^{eq}$ for the equilibrium regime of the random walk, which is reached after a sufficiently large number of jumps $n$.
  The Laplace transform for the random variable $t_{x}$ is given by
\begin{eqnarray}
E(\exp(-ut_{x}))&=&\prod_{i=1}^{n_{x}}\int_{0}^{\infty}d\tau_{i}\,\psi(\tau_{i})e^{-u\tau_{i}}\nonumber\\
&=&\exp\left(-nP_{x}^{eq}h(u)\right).
\label{TL}
\end{eqnarray}
Evidently, $h(u)$ is a monotonically increasing function for $u>0$, with $h(0)=0$.

In order to investigate the ergodic phase transitions for the trapping process, it
is interesting to introduce  the generating function $\hat{\rho}(\xi)$ developed in Ref. \cite{Godre} and successfully used in Ref. \cite{RB} for
describing time average observables. We will assume here, however, that
 the generating function depends on
  an additional parameter $\beta>0$ as follows
\begin{equation}
\hat{\rho}(\beta,\xi)=E\left(\left(\beta+\xi[\mathcal{O}]\right)^{-1}\right),
\label{GF}
\end{equation}
where
\begin{equation}
[\mathcal{O}]= \frac{\sum_{j=1}^{L}\mathcal{O}_{j}t_{j}}{\sum_{j=1}^{L}t_{j}},
\end{equation}
with the square bracket denoting,
  hereafter, the average of a given operator with respect to
the set $\left\{t_{1},\ldots,t_{L}\right\}$.
The generating function (\ref{GF})
 is related to the corresponding density function of time average observable $\bar{\mathcal{O}}$ as
\begin{eqnarray}
\rho(\bar{\mathcal{O}})&=&E\left( \delta\left(\bar{\mathcal{O}}-[\mathcal{O}]\right)\right)\nonumber\\
&=&-\frac{1}{\pi}\lim_{\epsilon\rightarrow0}\mbox{Im}\,\frac{1}{\bar{\mathcal{O}}+i\epsilon}\lim_{\beta\rightarrow1}\hat{\rho}\left(\beta,\frac{-1}{\bar{\mathcal{O}}+i\epsilon}\right).
\label{FTO}
\end{eqnarray}
Notice that $\xi$ must always be chosen in order to assure that $\beta+\xi [{\mathcal{O}}]>0$ for any set $\left\{t_{1},\ldots,t_{L}\right\}$.

\subsection{Ergodicity and weak nonergodicity}

Physically, we introduce the idea of a weakly nonergodic behavior  by demanding
that the lattice be
 not subdivided, from the dynamical point of view,
  in mutually inaccessible regions. This is equivalent to impose that \cite{RB}
\begin{equation}
0<P_{x}^{eq}<1\ {\rm for\ all\ } 1\leq x\leq L.
\label{irred}
\end{equation}
This condition is common to ergodic and weakly nonergodic systems.
 In fact,  
 we say that a CTRW is ergodic if  the condition (\ref{irred}) holds and,
  besides, the PDF (\ref{FTO}) could be written as $\rho(\bar{\mathcal{O}}) = \rho_E(\bar{\mathcal{O}})\equiv \delta(\bar{\mathcal{O}}-\left\langle \mathcal{O}\right\rangle)$, where
\begin{equation}
\left\langle \mathcal{O}\right\rangle=\sum_{j=1}^{L}P_{j}^{eq}\mathcal{O}_{j}
\end{equation}
will denote, hereafter,
  the equilibrium ensemble average, being $\mathcal{O}_{j}$   the value of the observable $\mathcal{O}$ when the particle is at the lattice point $j$. The PDF for the ergodic regime can be rewritten as
\begin{eqnarray}
\rho_{E}(\bar{\mathcal{O}})=-\frac{1}{\pi}\lim_{\epsilon\rightarrow0}\mbox{Im}\,\frac{1}{\bar{\mathcal{O}}+i\epsilon}\lim_{\beta\rightarrow1}\left(\beta-\frac{\left\langle \mathcal{O}\right\rangle}{\bar{\mathcal{O}}+i\epsilon}\right)^{-1}.
\label{Fdelta}
\end{eqnarray}
Combining Eqs. (\ref{FTO}) and (\ref{Fdelta}) we obtain the corresponding ergodic generating function:
\begin{eqnarray}
\hat{\rho}_{E}(\beta,\xi)=\frac{1}{\beta+\xi\left\langle \mathcal{O}\right\rangle}.
\label{ERGf}
\end{eqnarray}
The generating function
 (\ref{ERGf}) will be used hereafter as an unequivocal mark of ergodicity. In fact, we call weakly nonergodic a system 
which obeys (\ref{irred})  but 
 for which
 the generating function (\ref{FTO}) is not equivalent to (\ref{ERGf}). In other
 words, a weakly nonergodic system is a system where it is possible to get to
 any state from any state, but still  time and ensemble averages do  not
 coincide.

\section{Distribution densities of time averaged observables}
\label{sec3}

The generating function defined by Eq (\ref{GF}) can be rewritten in the following form
\begin{eqnarray}
\hat{\rho}(\beta,\xi)&=&\int_{0}^{\infty}ds\int_{0}^{\infty}dt\int_{0}^{\infty}dt_{1}\Psi(t_{1})\ldots\int_{0}^{\infty}dt_{L}\Psi(t_{L})\nonumber\\
&\times&\delta\left(t-\sum_{j=1}^{L}t_{j}\right)e^{-\left(\beta+\xi[\mathcal{O}]\right)s},
\label{fbe}
\end{eqnarray}
where $\Psi(t_{x})$ is the corresponding PDF for $t_{x}$.
We also demand
\begin{eqnarray}
\mathcal{U}_{j}=\beta+\xi\mathcal{O}_{j}>0,\qquad 1\leq j\leq L,
\label{uj}
\end{eqnarray}
in order to guarantee that $\beta+\xi [{\mathcal{O}}]>0$.
Now, by using the Laplace transformation given by Eq. (\ref{TL})
and taking into account that
\begin{eqnarray} \delta\left(t-\sum_{j=1}^{L}t_{j}\right)=\frac{1}{2\pi}\int_{-\infty}^{\infty}dk\,e^{ik\left(t-\sum_{j=1}^{L}t_{j}\right)},
\label{delta}
\end{eqnarray}
 Eq.
  (\ref{fbe}) can be cast, after successive changes of variables similar to those
  one performed in Ref. \cite{RB}, in the form
\begin{eqnarray}
\hat{\rho}(\beta,\xi)=-\int_{0}^{\infty}\int_{-\infty}^{\infty}dsdk\,\frac{e^{ -n\sum_{j=1}^{L}P_{j}^{eq}h(-ik+\xi\mathcal{O}_{j} s)}}{2\pi(k-i\beta s)^{2}}.\nonumber\\
\label{fbeC}
\end{eqnarray}
The $k$-integration   can
 be performed by using the Cauchy  formula, leading finally to
\begin{eqnarray}
\hat{\rho}(\mathcal{U}) =-\int_{0}^{\infty}ds\,\frac{1}{s}\frac{d}{d\beta}\,
e^{ -n\langle h(
 s\mathcal{U})\rangle }.
\label{MFBE}
\end{eqnarray}
Equation (\ref{MFBE}) is similar to that one found in Ref. \cite{RB} for the particular, but very important, case of one-sided L\'evy PDF, for which $h(u)=Cu^{\alpha}$, with $0<\alpha<1$. Here, however,
 we did not use either the generalized central limit theorem or scaling hypothesis.

 In order to solve equation  (\ref{MFBE}) for arbitrary functions $h(u)$,
 we introduce a new parametrization for the integral
\begin{eqnarray}
\mu(s) =\langle h(s\mathcal{U}) \rangle,
\label{inv1}
\end{eqnarray}
from which it is always possible to obtain $s=s(\mu)$ since
\begin{eqnarray}
\frac{d\mu}{ds}=
\langle \mathcal{U}h'(s\mathcal{U}) \rangle>0.
\label{inv2}
\end{eqnarray}
Notice that $\mu(0) = 0$ and $\mu(\infty) = \mu_{max}$.
With the new parametrization, Equation  (\ref{MFBE}) reads
\begin{eqnarray}
\hat{\rho}(\mathcal{U}) =n\int_{0}^{\mu_{max}}
f(\mu)
e^{-n\mu}\,d\mu ,
\label{xx}
\end{eqnarray}
where
\begin{eqnarray}
f(\mu) =
\frac{\langle h'(s(\mu)\mathcal{U}) \rangle}
{\langle \mathcal{U} h'(s(\mu)\mathcal{U}) \rangle}.
\label{xf}
\end{eqnarray}
The integral (\ref{xx}) can be evaluated in the
 limit $n\to\infty$, leading to (see Appendix \ref{Appendix} for details)
\begin{eqnarray}
\hat{\rho}(\mathcal{U}) =
\lim_{\mu\to 0^+} f(\mu) =
\lim_{s\rightarrow0^{+}}
\frac{\langle h'(s \mathcal{U}) \rangle}
{\langle \mathcal{U} h'(s \mathcal{U}) \rangle},
\label{xx1}
\end{eqnarray}
from which the PDF of time averaged observables $\rho(\bar{\mathcal{O}})$ can be obtained by using Eq. (\ref{FTO}), generalizing the previous results obtained in
\cite{RB}. Notice that Eq. (\ref{xx1}) depends only on the behavior of $h(u)$ near $u=0^{+}$, confirming  that only the asymptotic tail behavior of $\psi(\tau)$ is relevant in the limit $n\rightarrow\infty$. It should also be emphasized that the sojourn time $\tau$  is a non-negative unbounded random variable. Hence,
 only  functions $\exp(-h(u))$ that are Laplace transforms of PDF $\psi(\tau)$ with support on $[0,\infty)$ are relevant here.  According to Bernstein's
 theorem \cite{Feller}, the functions $\exp(-h(u))$ must be completely
 monotonic, {\em i.e.}, they should obey
\begin{equation}
(-1)^{k}\frac{d^{k}}{du^{k}}\exp(-h(u))\geq0,\qquad u>0.
\label{Ber}
\end{equation}

\section{Ergodic and Weakly Nonergodic Regimes}
\label{sec4}

The first conclusion that one can draw from Eq. (\ref{xx1}) is that systems with finite average trapping time $\bar{\tau}$ are ergodic since for such cases $\lim_{u\to 0^{+}} h'(u)=\bar{\tau}$, implying that Eq. (\ref{xx1}) reduces to Eq. (\ref{ERGf}).
On the other hand, the relevant weakly nonergodic models with diverging average trapping time are those ones considered in \cite{RB}, for which $h(u) = Cu^\alpha$, for small non negative $u$, with $0<\alpha<1$. The associate PDF in this case are the well known stable L\'evy densities, for which the asymptotic behavior
for large $\tau$ is given by
\begin{equation}
\psi(\tau) \sim A_{\alpha}\tau^{-(1+\alpha)}.
\label{levy}
\end{equation}
For such models, Eq. (\ref{xx1}) reads simply
\begin{eqnarray}
\hat{\rho}(\mathcal{U}) =
\frac{\langle   \mathcal{U}^{\alpha -1 }  \rangle}
{\langle \mathcal{U}^{\alpha   }\rangle},
\label{lamperti}
\end{eqnarray}
leading to a density function of Lamperti type \cite{RB}, where the ergodic regime
can be recovered  in the limit $\alpha\to 1^-$.
Interestingly enough, many other subtler   ergodic solutions do also exist.
This is the case, for instance, of the function
$h(u) = -Cu \log u$ for small $u$, for which
Eq. (\ref{xx1}) also reduces to Eq. (\ref{ERGf}). This case corresponds namely
to a PDF that asymptotically tends to the L\'evy density  (\ref{levy}) with $\alpha=1$.

Since  PDF with finite average sojourn  time $\bar{\tau}$ give rise to ergodic behavior, it would be worthy
to classify the possible PDF $\psi(\tau)$ with diverging $\bar{\tau}$
in order to identify possible nonergodic regimes.
  Since
one must demand $E(1) =1 $, all PDF $\psi(\tau)$ shall
asymptotically decrease faster than $\tau^{-1}$. On the other hand, in order
to have $\bar\tau = E(\tau)$ diverging, $\psi(\tau)$ cannot
 decrease  faster than $\tau^{-2}$.
Hence, a PDF $\psi(\tau)$ with diverging average sojourn  time $\bar{\tau}$
must obey
\begin{equation}
\frac{A_1}{\tau^2}  \le \psi(\tau) < \frac{A_0}{\tau},
\label{interval}
\end{equation}
for large $\tau$, with $A_0$ and $A_1$ arbitrary positive constants.
The cental point here is that the class of  power-like functions like $h(u) = Cu^\alpha$, with $0<\alpha<1$,
  which correspond to the Levy distributions
(\ref{levy}), does not exhaust all the interval (\ref{interval}).
In particular, the lower bound  of the interval does  not correspond to
any of these functions. As we have already shown, such case
(ergodic and with $\alpha =1$)
corresponds indeed to the function
$h(u) = -Cu \log u$ for small $u$.  PDF with asymptotic behavior
of the type
\begin{equation}
\psi(\tau) \sim \frac{B_\gamma}{\tau\log^\gamma \tau},
\label{upper}
\end{equation}
where $B_\gamma$ and $\gamma > 1$ are constants,  obey (\ref{interval}). They,
  in fact, accumulate in the upper bound. On the other hand,
  PDF of the type
\begin{equation}
\psi(\tau) \sim C_\nu\frac{\log^\nu \tau}{\tau^2},
\label{lower}
\end{equation}
where $C_{\nu}$ and $\nu > 0$ are constants,
also belong  to the interval and accumulate
in the lower bound.

In order to extend our analysis to consider the
PDF with logarithmic terms  are those ones of (\ref{upper}) and
(\ref{lower}), one can make use of
Karamata's Abelian and Tauberian theorems \cite{Feller}  for the Laplace-Stieltjes transform
\begin{equation}
  e^{-h(u)} = \int_0^\infty e^{-u\tau } d{\Upsilon}\left(\tau\right),
\end{equation}
for $u>0$, which states that
\begin{equation}
\label{karamata1}
 e^{-h(u)} \sim \Gamma(\rho+1){\Upsilon}\left( \frac{1}{u}\right)
\end{equation}
for $u\to 0+ $,
if
\begin{equation}
\label{karamata2}
\lim_{\tau\to\infty}
\frac{{\Upsilon}\left( \tau x \right)}{{\Upsilon}\left( \tau\right)}\to x^\rho.
\end{equation}
In the present case, since
 ${\Upsilon}\left( \tau\right)$ is the cumulative distribution function  associated to $\psi(\tau)$, the condition (\ref{karamata2}) is automatically fulfilled  with $\rho=0$. Hence, from (\ref{karamata1}), we have
 \begin{equation}
 \label{karamata3}
    h(u)  \sim  1 - {\Upsilon}\left( \frac{1}{u}\right)
 \end{equation}
 for $u\to 0+ $.

 For the PDF
(\ref{upper}), we have, by using (\ref{karamata3})
\begin{equation}
h(u) \sim \frac{B_\gamma}{\gamma-1} \frac{1}{|\log u|^{\gamma-1}},
\end{equation}
for small $u$ and
$\gamma > 1$. For this case, we have
$h'(u) =   B_\gamma/(u|\log u|^{\gamma})$ for small $u$, implying, from (\ref{xx1}),
that
\begin{equation}
\hat\rho(\mathcal{U}) = \langle\mathcal{U}^{-1} \rangle,
\end{equation}
for any value of $\gamma>1$,
which, interestingly, coincides with  the limit $\alpha\to 0$ of the
Lampertian case given by (\ref{lamperti}).  For the PDF (\ref{lower}),
on the other hand, we have
\begin{equation}
h(u) \sim C_\nu u|\log u|^\nu,
\end{equation}
for small $u$ and
$\nu > 0$, which give rises to a ergodic generating function
\begin{equation}
\hat\rho(\mathcal{U}) = \langle\mathcal{U} \rangle^{-1},
\end{equation}
for
any $\nu > 0$, which, of course, also coincides  with the
limit $\alpha\to 1$ of the
Lampertian case   (\ref{lamperti}). The Lampertian
generating function (\ref{lamperti}), with
$0\le \alpha \le 1$, seems to be enough to describe any
CTRW of the type considered up to here.

\section{Stable PDF}
\label{sec5}

Any physical application   of the preceding sections results
would  require, of course, stable PDF $\psi(\tau)$. We recall that a
stable PDF $S_{\alpha}(\tau;\sigma,\beta,\mu)$
is characterized by four parameters: $0<\alpha\leq2$,
which determines the asymptotic falling tails;
$\sigma\geq0$ is the corresponding scale; $-1\leq\beta\leq1$ being
the skewness; and $\mu$  the shift parameter.
 The  support of a generic stable density is given by \cite{Taqqu}
\begin{equation}
\left\{\begin{array}{ll}
      [\mu,\infty),\,\,0<\alpha<1,\beta=1,\\
      (-\infty,\mu],\,\,0<\alpha<1,\beta=-1,\\
      (-\infty,+\infty),\,\,\mbox{otherwise}.
                 \end{array}
                 \right. \label{supp}
\end{equation}
Since $\tau$ is an unbounded non-negative random variable we disregard
the case $\beta=-1$. The natural stable PDF in this case are
\begin{equation}
\psi(\tau)=S_{\alpha}(\tau;\sigma,1,0),
\label{mod1}
\end{equation}
with $0< \alpha < 1$, which have the asymptotic behavior given by
(\ref{levy}).
There are, however, several  other useful possibilities.

Let us
 redefine the support of a stable PDF by taking, first, $\mu=C>0$ and then  $\tau\rightarrow|\tau-C|$ so that, for $\beta\neq-1$, one has
 \begin{equation}
\psi(\tau)= \left\{
 \begin{array}{cl}
      N S_{\alpha}(|\tau-C|;\sigma,\beta,0), & {\rm for\ }\tau\ge 0, \\
      0, & {\rm for\ }\tau <  0,
      \end{array}
      \right.
        \label{pstab}
\end{equation}
where $N$ is the pertinent normalization constant
and $0<\alpha<2$.
 The general asymptotic behavior  of the PDF     (\ref{pstab})
 is given by (\ref{levy}) \cite{Taqqu}. The average sojourn time $\bar{\tau}=E(\tau)$ diverges as $\tau^{1-\alpha}$ for $0<\alpha<1$, and logarithmically for $\alpha=1$. According to
the results of the previous section,
 the cases for which $\bar\tau$ is finite are ergodic.
 Some explicit examples will help to illustrate our main results.

 \subsection{A non-negative delta sequence PDF}

Consider the non-negative delta sequence PDF on $[0,\infty)$ given by
\begin{equation}
\psi(\tau)=N \frac{\sin^2 \gamma(\tau-C) }{\gamma(\tau-C)^2}
,\label{WB1} \end{equation}
where $N$ is the appropriate  normalization constant and $\gamma$
and   $C$ are positive parameters.  PDF (\ref{WB1}) is the
  one-sided version of the density defined in Ref. \cite{Walter}.
 In the limit $\gamma\to\infty$, it corresponds to a delta function centered
 in $\tau=C$.
   The function $h(u)$ associated to (\ref{WB1}) is given, for small non-negative $u$, by
\begin{equation}
h(u)  = \left(D-N\log\left(\frac{u}{2\gamma}\right)\right)\frac{u}{2\gamma}+
O\left(\left(\frac{u}{2\gamma} \right)^{2}\right),
\end{equation}
where $D$ is a constant depending on $\gamma$ and $C$. Despite of having
a diverging first moment $\bar\tau$, the PDF (\ref{WB1}) gives rise to
an ergodic regime since the dominant term in the function $h(u)$  for small $u$
is precisely $-u\log u$, that it is known from the results of the last
 section to reduce (\ref{xx1}) to the
ergodic generating function (\ref{ERGf}).

\subsection{The Cauchy stable PDF}

 The Cauchy stable PDF, defined  for  $\tau\in[0,\infty)$ as
\begin{equation}
\psi(\tau)=N\frac{\sigma}{(\tau-C)^{2}+\sigma^{2}},
\end{equation}
where $N$ is the appropriate  normalization constant and $\sigma$
and   $C$ are positive parameters, is another
example of a  PDF with diverging $\bar\tau$ but that, nevertheless,
gives rise to an ergodic regime. The function $h(u)$ in this case
is given by
\begin{equation}
h(u)=\left(D-N\log (\sigma u)\right)\sigma u+O((\sigma u)^{2}),
\end{equation}
for small and non-negative $u$, where $D$ is a constant depending
on $C$ and $\sigma$. Again, we have a situation where the logarithmic term
dominates and renders (\ref{xx1}) to the ergodic expression (\ref{ERGf}).

\subsection{The L\'evy stable PDF}
Based on the one-sided PDF (\ref{pstab}), let us consider the
$(\alpha=1/2)$
 L\'evy density
 with support on $[0,\infty)$ given by
\begin{equation}
\psi(\tau)=N\sqrt{\frac{\sigma}{2\pi}}\,\frac{\exp \left( \frac{-\sigma}{2|\tau-C|}\right)}{|\tau-C|^{3/2}},\label{Lev1}
\end{equation}
where $N$ is the appropriate  normalization constant and $\sigma$
and   $C$ are positive parameters. The  corresponding  $h(u)$  function can be calculated for $u\rightarrow0^{+}$ as
\begin{equation}
h(u)=N \sqrt{2\sigma u}+(C-D) u+O((\sigma u)^{3/2}),\label{hul}
\end{equation}
where
\begin{equation}
D = 2N\left(\sqrt{\frac{\sigma C}{2\pi}}e^{-\frac{\sigma}{2C}} + (1-N)\sigma \right).
\end{equation}
For large times, the term $ (\sigma u)^{1/2}$ prevails, leading to the $(\alpha=1/2)$ Lamperti's statistics (\ref{lamperti}). This
 result coincides with that one obtained in Ref. \cite{RB} for the PDF (\ref{mod1}). Note, however, that $\sigma\rightarrow0$ leads to an ergodic regime, regardless of having
 $\alpha=1/2$!
 This is a somewhat surprising result.

\section{Non-identical trapping time PDF}
\label{sec6}

So far we have considered  only situations where a single trapping time PDF $\psi(\tau)$ is sufficient to describe  the random dynamics.
However, there are situations  where the trapping times  are not identically distributed over the lattice points.
This is the case, for instance, of the  fluorescence blinking observed in some colloidal nanocrystals, namely the case of certain quantum dots systems \cite{BHMDBD,MB}. Once  a laser pulse incides on a quantum dot of these systems,
its fluorescence intensity   randomly switches
between bright (on)  and dark (off) states. The blinking quantum dots are typically characterized by means of the statistics of {on}/{off} times, whose distributions exhibit power law decay $\psi_{k}(\tau)\sim A_{k}\tau^{-(1+\alpha_{k})}$, where $k$ corresponds to the {on}/{off}  states and $0<\alpha_{k}<1$. Nonergodicity for such systems has been reported from experimental observations \cite{BHMDBD}.

Our approach can be generalized to include such kind of situation.
To this end, let us
consider now a CTRW model described by a set of arbitrary trapping time PDF $\psi_{j}(\tau)$ governing   the sojourn times $\tau_j$ of the $L$ lattice points $1\leq j\leq L$. In such non-homogeneous lattice, the sojourn times  are not identically distributed
anymore, but they
are still independent random variables. We can extend all the results Sections \ref{sec2} and \ref{sec3} essentially by replacing $h(u)$
 by $h_{j}(u)$, resulting finally in
 \begin{eqnarray}
\hat{\rho}(\mathcal{U})=\lim_{s\rightarrow0^{+}}\frac{\sum_{j}^{L}P_{j}^{eq}h'_{j}(s\mathcal{U}_{j})}{\sum_{j}^{L}P_{j}^{eq}\mathcal{U}_{j}h'_{j}(s\mathcal{U}_{j})},
\label{nittd}
\end{eqnarray}
instead of (\ref{xx1}).
 The first conclusion we get from (\ref{nittd}) is
 that weak nonergodicity is also present
 in this case, since we might have nonergodic regimes for which
 $0<P_{j}^{eq}<1$ for all $1\leq j\leq L$, assuring that the lattice
 is not dynamically subdivided in  mutually inaccessible regions. Furthermore,
 we have also that  the weak ergodicity breaking is not a structurally stable phenomenon since ergodic transitions are impossible for systems where  the sojourn times $\tau$ are not identically distributed. Even for systems where the average values $\bar{\tau}_{j}$ are finite for all  lattice points,   but different, one has the predominance of nonergodicity. In order to  illustrate this scenario, some explicit examples are useful. For any of the three cases depicted in
 Table \ref{table},
 \begin{table}
\begin{tabular}{| c | c |  c |  }
  \hline
  $h_j(u)$ & $\psi_j(\tau)$ & $\bar\tau_j$ \\
  \hline
$C_ju$  & faster than $\tau^{- 2 }$ &  finite \\
\hline
$-C_ju \log u $  & $A_j\tau^{- 2 }$ & $\log\tau$ \\
 \hline
$C_ju^\alpha$    & $A_j\tau^{-(1+\alpha)}$&  $\tau^{1-\alpha}$ \\
      \hline
\end{tabular}
\caption{Some possible non-homogeneous PDF over the lattice. The first column depicts the behavior of $h_j(u)$ for small non-negative $u$,
 corresponding to the lattice point $j$.
 The second column has the associate PDF asymptotic behavior for large $\tau$.
 The last column shows how the average sojourn time diverges.
 We assume that
$C_j$ are positive constants and $0<\alpha<1$.}
  \label{table}
 \end{table}
we have that the expression for $\rho(\bar{\mathcal{O}})$ as defined by
 (\ref{FTO}) are obtained from the corresponding single trapping time PDF cases by replacing $P_{j}^{eq}\rightarrow P_{j}^{eq}C_{j}$, {\em i.e.}, by
 replacing
\begin{equation}
\left\langle\mathcal{O} \right\rangle\rightarrow\frac{\left\langle C\mathcal{O}\right\rangle}{\left\langle C\right\rangle},\label{none}
\end{equation}
what prevents
any transition between  weakly nonergodic and ergodic regimes. We can also conclude that the arising of narrow PDF does not necessarily mean that a system has reached the ergodic regime.

Another interesting dynamical aspect of the large time
 evolution of systems with non-homogeneous
 PDF is the competition  between the lattice points. Let us introduce
  the survival rate $\eta_{kl}$ between  two lattice points $k$ and $l$, defined as
\begin{eqnarray}
\eta_{kl}=\lim_{u\rightarrow0^{+}}\frac{h'_{k}(u)}{h'_{l}(u)}.
\label{srate}
\end{eqnarray}
The survival rate is, of course, related to the ratio
$ \bar\tau_k /\bar\tau_l $
for the lattices points $k$ and $l$. Such relation, however, is not
obvious for general PDF. Notice that
\begin{equation}
\label{diflaplace}
e^{-h_i(u)}h_i'(u) = \int_0^\infty e^{-u\tau} d\Lambda_i(\tau),
\end{equation}
where
\begin{equation}
\Lambda_i(\tau)=\int_0^\tau s\psi_i(s)\, ds.
\end{equation}
Provided that $\Lambda_i(\tau)$ obey a condition like  (\ref{karamata2}),
we have by using Karamata's theorem for (\ref{diflaplace})
\begin{equation}
h_i'(u) \sim \Gamma(\rho_i +1) {\Lambda_i\left(\frac{1}{u}\right)} ,
\end{equation}
for $u\to 0^+.$
Since $\Lambda_i(\infty) = \bar\tau_i$, we conclude that the survival
rate (\ref{srate}) does no coincide, in general,  with the ratio $ \bar\tau_k /\bar\tau_l $. Nevertheless, the survival rate is indeed an apropriate
 quantity to compare the residence time of the CRWT dynamics
in different points of the non-homogeneous lattice.

According to (\ref{srate}), two types of behavior can occur.
Both states coexist if $\eta_{kl}$ is finite and non-negligible.
If $\eta_{kl}$ is infinite or vanishes, then, respectively,
 state $k$ prevails over $l$, or vice versa.
 In the case where prevailing states exist, only them are relevant for the calculation of the density of time averages.
 Notice that even if a state $k$ has a visitation fraction negligible when compared to another state $l$, {\em i.e.} $P_{k}^{eq}\ll P_{l}^{eq}$, state $k$ can still prevail  over state $l$ if $\eta_{kl}=\infty$.
Irrelevant states can be visited many times, but the time spent among them
 by the CTRW dynamics
 is negligible and, hence, they should not contribute to the time average.
 This is the  mechanism behind  the  nonergodicity of
  non-homogeneous lattices.

\section{Concluding remarks}
\label{sec7}

Weak ergodicity breaking has been investigated \cite{RB,BB,HBMB} by considering Levy PDF (\ref{levy}), for which the time-average densities are given by a Lamperti distribution by means of (\ref{lamperti}). For this class of PDF, the weakly nonergodic regime breaks up, giving origin to an ergodic one, in the limit $\alpha\to 1$. Despite that our
results strongly indicate that Lamperti distributions seems to be enough to describe any
time-average density for
CTRW with PDF obeying (\ref{interval}), our results also show that the limit
$\alpha\to 1$ is not the only way to obtain ergodic regimes.
This fact is explicitly illustrated by the PDF given by (\ref{Lev1}),
for which an ergodic regime arises for  $\sigma\rightarrow0$,
irrespective of having $\alpha=1/2$ for such PDF.
This results challenges the naive association between
 ergodicity/weak nonergodicity and diffusion/subdiffusion
 as suggested in   \cite{HBMB}, which indeed
appears reasonable at first sight since models with PDF (\ref{levy})
typically
exhibit  anomalous diffusion characterized by
 $\left\langle x^{2}\right\rangle\sim t^{\alpha}$. The characterization
 of any possible universal behavior close to the weak ergodic breaking
 cannot by
 achieved by analyzing only the breaking associated with   $\alpha\to 1$.

As to the case of non-identical sojourn times over the lattice, the issue of weak ergodicity breaking is still more involved. Even for PDF with finite average sojourn times $\bar\tau_j$, we do not have an ergodic regime, in contrast with the
case of homogeneous lattices, for which a finite $\bar\tau$ will necessarily imply in an ergodic regime. This nonergodic regime can be considered also as weakly nonergodic since, from
(\ref{none}), we have that
the phase space is similar to the homogeneous case being, in particular,
  not subdivided
  in  mutually inaccessible regions. Ergodicity, in this case, is recovered
  in the limit of equal PDF. This can be illustrated by a simple explicit example. Let us consider a Markov chain with only two states $A$ and $B$ and with the transition matrix
  \begin{equation}
  W = \frac{1}{2}\left(\begin{array}{cc}1 & 1\\
  1 & 1
  \end{array} \right).
  \end{equation}
  The corresponding invariant weights are
 $P^{eq}_A = P^{eq}_B = 1/2$. Assume now that the sojourn times $\tau_A$ and
 $\tau_B$ are governed by the PDFs
 $\psi_A(\tau) = \lambda_Ae^{-\lambda_A\tau }$ and
$\psi_B(\tau) = \lambda_Be^{-\lambda_B\tau }$. After $n \gg 1$ steps, we will have $n_A\approx n_B \approx n/2$, but the visitation fraction of the dynamics
in the two states  will be in general different since
\begin{equation}
\frac{n_AE(\tau_A)}{n_BE(\tau_B)} \approx \frac{\lambda_B}{\lambda_A}.
\end{equation}
For the case of diverging $\bar\tau_j$, besides
  of the consequences associated with (\ref{none}), we have also new features associated with the possible divergence or vanishing of the survival rate (\ref{srate}). This leads to the possibility of a new weakly nonergodic regime, for which, despite of the CTRW spreading over all lattice points, the time-averaged distributions will depend only on some points, namely the prevailing ones according to the survival rate (\ref{srate}). Despite these points are not surprising from the probabilistic point of view, they
  certainly deserve a deeper physical investigation.

\acknowledgments

The authors gratefully acknowledge stimulating discussions with E. Barkai, C.J.A. Pires, and R.D. Vilela.
The authors wish to recognize, in particular, an anonymous referee for suggesting the explicit example of Section VII and several other improvements in the paper.
This work was supported by the Brazilian agencies CNPq and FAPESP.

\appendix

\section{}
\label{Appendix}
The first observation about the limit  of large $n$ of equation (\ref{xx}) is that one cannot apply Watson's lemma \cite{BH} directly since we cannot assure, a priori, if
the derivatives of $f(\mu)$ are finite or not for $\mu\to 0^+$. Let us take,
for instance, the first derivative
\begin{equation}
\lim_{\mu\to 0^+}f'(\mu) =
\lim_{s\to 0^+}\left(
\frac{\langle \mathcal{U}h''(s \mathcal{U}) \rangle}
{\langle \mathcal{U}h'(s \mathcal{U}) \rangle^2} -
\langle  h' (s \mathcal{U}) \rangle
\frac{\langle\mathcal{U}^2h''(s \mathcal{U}) \rangle}
{ \langle \mathcal{U}h'(s \mathcal{U}) \rangle^3  }\right)
\label{limit}
\end{equation}
where (\ref{inv2}) was used. Although, apparently, for all physically
relevant functions $h(u)$ the limit (\ref{limit}) is indeed finite, one
cannot rule out, in principle,
 possible situations where it might diverge. Fortunately,
thanks to the boundedness of $f(\mu)$, one can evaluate (\ref{xx}) without
using Watson's lemma.

By introducing the new variable $v = n\mu$, Equation (\ref{xx})  can be cast in the
limit $n\to \infty$ as
\begin{equation}
\hat{\rho}= \lim_{n\to\infty} \int_{0}^{n\mu_{max}}
f\left(\frac{v}{n} \right)
e^{-v}\, d v,
\label{xxAp}
\end{equation}
with
  $f(\mu)\ge 0$   given by (\ref{xf}). First of all, let us suppose that
  $\mu_{max}<\infty$.
   Equation (\ref{xxAp}) can be decomposed in this case as
\begin{equation}
\hat\rho = \lim_{n\to\infty}  I_0(n) + \lim_{n\to\infty} I_1(n) ,
\end{equation}
where
\begin{equation}
I_0(n) = \int_0^{\sqrt{n}\mu_{max}} f\left(\frac{v}{n} \right)
e^{-v}\, d v
\end{equation}
and
\begin{equation}
I_1(n) = \int_{\sqrt{n}\mu_{max}}^{n\mu_{max}} f\left(\frac{v}{n} \right)
e^{-v}\, d v.
\end{equation}
For both integrals, we have
\begin{equation}
f^-_ig_i(n)\le I_i(n) \le
f^+_ig_i(n),
\label{bounds}
\end{equation}
with $i=0,1$,
where
\begin{eqnarray}
g_0(n) &=& 1 - e^{-\sqrt{n}\mu_{max}}, \\
g_1(n) &=&  e^{-\sqrt{n}\mu_{max}} - e^{- {n}\mu_{max}},  \\
f^-_i &=&  \inf_{\mu\in \mathcal{I}_i} f(\mu),\\
f^+_i &=&  \sup_{\mu\in \mathcal{I}_i} f(\mu),
\end{eqnarray}
where $\mathcal{I}_0 = \left[0,\frac{\mu_{max}}{\sqrt{n}} \right]$ and
$\mathcal{I}_1 = \left[\frac{\mu_{max}}{\sqrt{n}} , \mu_{max} \right]$.
Since $f(\mu)$ is  bounded, we have $f_1^- \le f_1^+ <\infty$. Taking into
account that
 $\lim_{n\to\infty} g_1(n) = 0$, one has from (\ref{bounds}) that  $\lim_{n\to\infty} I_1(n) = 0$. As to the integral $I_0(n)$, notice that    $\lim_{n\to\infty} f_0^- = \lim_{n\to\infty} f_0^+ =
 \lim_{\mu\to 0^+} f(\mu)$ and $\lim_{n\to\infty} g_0(n) = 1$, implying finally from
 (\ref{bounds}) that
 \begin{equation}
 \hat{\rho}= \lim_{n\to\infty} n\int_{0}^{ \mu_{max}}
f\left(\mu \right)
e^{-n\mu}\, d\mu = \lim_{\mu\to 0^+} f(\mu).
\label{final}
 \end{equation}
 The cases for which $\mu_{max}$ diverges can be treated in an analogous way,
 by choosing $\sqrt{n}\mu_*$  with finite $\mu_*$, instead of $\sqrt{n}\mu_{max}$, for the decomposition of (\ref{xxAp}) into $I_0(n)$ and $I_1(n)$ integrals, leading to the same final result (\ref{final}).

\end{document}